\documentclass[conference]{IEEEtran}
\IEEEoverridecommandlockouts
\usepackage{cite}
\usepackage{newtxtext,newtxmath}
\usepackage{algorithmic}
\usepackage{graphicx}
\usepackage{textcomp}
\usepackage{xcolor}
\usepackage{multirow}
\usepackage{multicol}
\usepackage{url}
\usepackage{stfloats}

\def\BibTeX{{\rm B\kern-.05em{\sc i\kern-.025em b}\kern-.08em
    T\kern-.1667em\lower.7ex\hbox{E}\kern-.125emX}}
\begin{document}

\title{Toward Less Hidden Cost of Code Completion with Acceptance and Ranking Models\\
}

\makeatletter
\newcommand{\linebreakand}{%
  \end{@IEEEauthorhalign}
  \hfill\mbox{}\par
  \mbox{}\hfill\begin{@IEEEauthorhalign}
}
\makeatother

\author{
\IEEEauthorblockN{1\textsuperscript{st} Jingxuan Li}
\IEEEauthorblockA{
\textit{Huawei Technologies Co., Ltd}\\
Shenzhen, China \\
lijingxuan1@huawei.com}
\and
\IEEEauthorblockN{2\textsuperscript{nd} Rui Huang}
\IEEEauthorblockA{
\textit{Huawei Technologies Co., Ltd}\\
Shenzhen, China \\
huangrui27@huawei.com}
\and
\IEEEauthorblockN{3\textsuperscript{rd} Wei Li}
\IEEEauthorblockA{
\textit{Huawei Technologies Co., Ltd}\\
Shenzhen, China \\
liwei563@huawei.com}
\linebreakand 
\IEEEauthorblockN{4\textsuperscript{th} Kai Yao}
\IEEEauthorblockA{
\textit{Huawei Technologies Co., Ltd}\\
Shenzhen, China \\
yaokai14@huawei.com}
\vspace{-0.5cm}
\and
\IEEEauthorblockN{5\textsuperscript{th} Weiguo Tan}
\IEEEauthorblockA{
\textit{Huawei Technologies Co., Ltd}\\
Shenzhen, China \\
tanweiguo@huawei.com}
\vspace{-0.5cm}
}

\maketitle
\IEEEpeerreviewmaketitle
\begin{abstract}
Code completion is widely used by software developers to provide coding suggestions given a partially written code snippet. Apart from the traditional code completion methods, which only support single token completion at minimal positions, recent studies show the ability to provide longer code completion at more flexible positions. However, such frequently triggered and longer completion results reduce the overall precision as they generate more invalid results. Moreover, different studies are mostly incompatible with each other. Thus, it is vital to develop an ensemble framework that can combine results from multiple models to draw merits and offset defects of each model.

This paper conducts a coding simulation to collect data from code context and different code completion models and then apply the data in two tasks. First, we introduce an acceptance model which can dynamically control whether to display completion results to the developer. It uses simulation features to predict whether correct results exist in the output of these models. Our best model reduces the percentage of false-positive completion from 55.09\% to 17.44\%. Second, we design a fusion ranking scheme that can automatically identify the priority of the completion results and reorder the candidates from multiple code completion models. This scheme is flexible in dealing with various models, regardless of the type or the length of their completion results. We integrate this ranking scheme with two frequency models and a GPT-2 styled language model, along with the acceptance model to yield 27.80\% and 37.64\% increase in TOP1 and TOP5 accuracy, respectively. In addition, we propose a new code completion evaluation metric, Benefit-Cost Ratio(BCR), taking into account the benefit of keystrokes saving and hidden cost of completion list browsing, which is closer to real coder experience scenario.

\end{abstract}

\begin{IEEEkeywords}
Code completion, neural networks, acceptance model, ranking model, evaluation metrics
\end{IEEEkeywords}

\section{Introduction} 
\label{section:Introduction}

Research shows that code completion is one of the most frequently used functions in the IDEs\cite{murphy2006java}. Code completion tools such as IntelliSense was first released as the main feature of Visual Basic 5.0 in 1996 and activated by default in the Visual Studio 2005 installation package. In recent years, the development of machine learning raises interest in research of the intelligent code completion area. Microsoft and Kite successively released IntelliCode\cite{IntelliCode} and Intelligent Snippets\cite{Kite-web} as code completion plug-ins for multi IDEs, respectively. TabNine\cite{Tabnine} also developed a code completion plug-in based on a deep learning model, which supports over 30 programming languages. When the developer writes code, the code completion engine can infer the potential following code fragments at the cursor position based on the written code context. Thus, it helps developer to improve the efficiency of coding by decreasing typing costs.

Traditional code completion methods embedded in IDEs rely on compile-time type information and specific grammatical rules to predict next tokens\cite{murphy2006java,li2017code}, which are costly and could not capture human's programming patterns well. In addition, they only trigger at the limited position with the completion results sorted in the alphabetic order. To address these problems, the concept of intelligent code completion was proposed\cite{bruch2009learning}. With the naturalness of the programming languages had been proved\cite{hindle2016naturalness}, researchers started to apply various learning-based algorithms and train models to learn code characteristics from large-scale codebases. At an early stage, statistical language model\cite{hindle2016naturalness}, such as N-gram, is a promising approach for code completion tools\cite{hou2010towards,hellendoorn2017deep}. \cite{5431761} proposed a Hidden Markov Model to complete multiple tokens at a time with abbreviated input. With the development of deep learning, researchers started to apply Recurrent Neural Networks (RNNs) styled models in source code modelling\cite{yang2019improve,svyatkovskiy2019pythia,aye2020sequence,li2017code} in the last five years. However, the performance of RNNs-styled models is limited by their vanishing/exploding gradients and sequential processing mechanism\cite{vaswani2017attention}. Most recently, transformers styled pretraining language models have overtaken RNNs and shown their advantages by achieving SOTA results in natural language processing (NLP) and natural language understanding (NLU) tasks\cite{devlin2018bert,radford2018improving,liu2019roberta,dai2019transformer}. \cite{liu2020self} transfers source code into Abstract Syntactic Tree (AST) based node sequence and predict type and value of next node at the same time with Transformer-XL. \cite{svyatkovskiy2020intellicode} introduces IntelliCode based on GPT-2 models to provide up to entire line completion. However, with all the progress, code completion is not without drawbacks as a system.

\textbf{a) Displaying too many false-positive suggestions.} Intelligent code completion algorithms can provide completion results at almost arbitrary positions within the code. Thus, the completion display frequency of these intelligent code completion algorithms increases a lot when the developer is typing code. Such frequently display results can improve coding efficiency by increasing the recall of completion. Meanwhile, it might also cause a drop of precision as too much false-positive suggestion is presented which reduces the coder experience. In general, user experience plays a more important role when developer choosing plug-ins, so it is vital to block the display of completion results when they are invalid.

\textbf{b) Lack of fusion ranking methods to combine results from multi code completion algorithms.} Due to the different interests of research, such as the different semantic representation of programming language\cite{alon2018general,brockschmidt2018generative,alon2018code2seq} and the different completion targets (API calls\cite{gu2016deep,fowkes2016parameter,nguyen2017exploring,svyatkovskiy2019pythia,svyatkovskoy2020fast}, AST node\cite{bielik2016phog,li2017code}, entire line\cite{svyatkovskiy2020intellicode}), each code completion algorithm has its distinct characteristics. In addition, the traditional code completion also has a different type of results sorted in alphabetical order. Current ensemble methods usually sort the candidates by the priority of the strategy. With all these comprehensive traditional and intelligent completion results, there is a lack of effective methods to integrate them, complement each other, and maximize completion efficiency. Thus, there is still a large room for completion result optimization, including the number of items listed in the completion result and the order of the items in the completion list.

\textbf{c) Lack of comprehensive assessment method to evaluate efficiency, including the hidden cost, of code completion tool.} Most tasks are to mask either code tokens or AST nodes in the code file according to certain rules, then using context to predict them. Since the targets of different researches are diverse, they proposed various evaluation metrics. For instance, top n accuracy and mean reciprocal rank (MRR) is commonly used for single token/AST node completion\cite{svyatkovskiy2019pythia,yang2019improve}, while BLEU-4 and edit similarity are used for multi tokens/full line completion\cite{svyatkovskiy2020intellicode}. Such various targets of code completion task make it difficult to compare their overall effects. Moreover, there is a gap between these offline evaluation metrics of the code completion model and the user experience in real scenarios. When using the code completion plug-in, it is impossible to know which part of the code should mask in advance. The accuracy evaluation metrics based on specific locations cannot fully reflect the actual use effect of the code completion tools. It ignores the time cost for developers to check the completion list and the negative impact of frequent invalid lists\cite{jin2018hidden}. This increases the difficulty of promoting the intelligent code completion research achievements in real code completion plug-ins. Therefore, a general method to evaluate the code completion efficiency without adding specific constraints is necessary.

To address the challenges mentioned above, we have conducted extensive investigation and simulation to collect data from code context and different code completion strategies(cf. Section \ref{section:Dataset} and \ref{section:Simulation data Generation}). Then, we apply the data to optimize the code completion tool from the perspective of making it more practical and user-friendly to developers. The following are our three main contributions:

\textbf{First, we introduce and train an acceptance model which can dynamically control whether to accept the completion results and display them to the developer}. This model uses features extracted from code context and the features of the code completion results from different strategies as input and outputs the probability that whether a correct result is in the aggregate completion list(cf. Section \ref{section:Acceptance of completion list}). 

\textbf{Second, we design and construct a fusion ranking scheme that can automatically identify the priority of the completion results and reorder the candidates provided by different completion strategies.} This scheme is flexible in dealing with various code completion strategies, regardless of the type or the length of their completion results (cf. Section \ref{section:Fusion ranking method}). In addition, our proposed scheme does not involve complex models, and the additional calculation and time-consuming are within a tolerable range. And it is easy to extend to new code completion algorithms and new development languages.

\textbf{Third, we design and implement a code completion evaluation method that is closer to real coder experience scenario.} It not only considers many factors such as keystrokes saving and completion list browsing, but also can be universally applied to the evaluation of different completion strategies (cf. Section \ref{section:Evaluation Method}). 

Experiments have proved that after using the acceptance and fusion ranking models, even a simple frequency-style code completion model superimposed on a GPT-2 styled model can bring good benefits (cf. Section \ref{section:Experiment Setup} and \ref{section:Result}). At the end, we conclude our study and mention future work (cf. Section \ref{section:Conclusion}).

\section{Motivation example} 
\label{section:Motivation}
Fig. \ref{fig:example2} shows two examples to illustrate the motivation of our code completion scheme. Fig. \ref{fig:example2}(a) presents a code completion system that integrates two different completion strategies, Global Frequency and Local Frequency. These two strategies provide completion candidates with their corresponding scores in the table at the given cursor location. It should be noted that different strategies might provide scores in different scales and dimensions. However, these candidates are all incorrect. Thus, the acceptance model should evaluate the confidence of all the candidates and reject the display of the completion list.
In Fig. \ref{fig:example2}(b), we integrate one completion strategy based on the deep learning model (GPT-2) together with two frequency strategies in the code completion system. The upper table presents a list of the completion candidates and scores provided by each strategy. As the correct results are involved in the candidate list, the acceptance model accepts the completion results and forwards the completion results to the fusion ranking model. The fusion ranking model ranks the completion candidates by their expected benefit shown in the lower table. This expected benefit is evaluated by the accuracy confidence and the token length of each completion candidate.

\begin{figure}[tbp]
  \centering
  \setlength{\abovecaptionskip}{-0.2cm}
  \includegraphics[width=\linewidth]{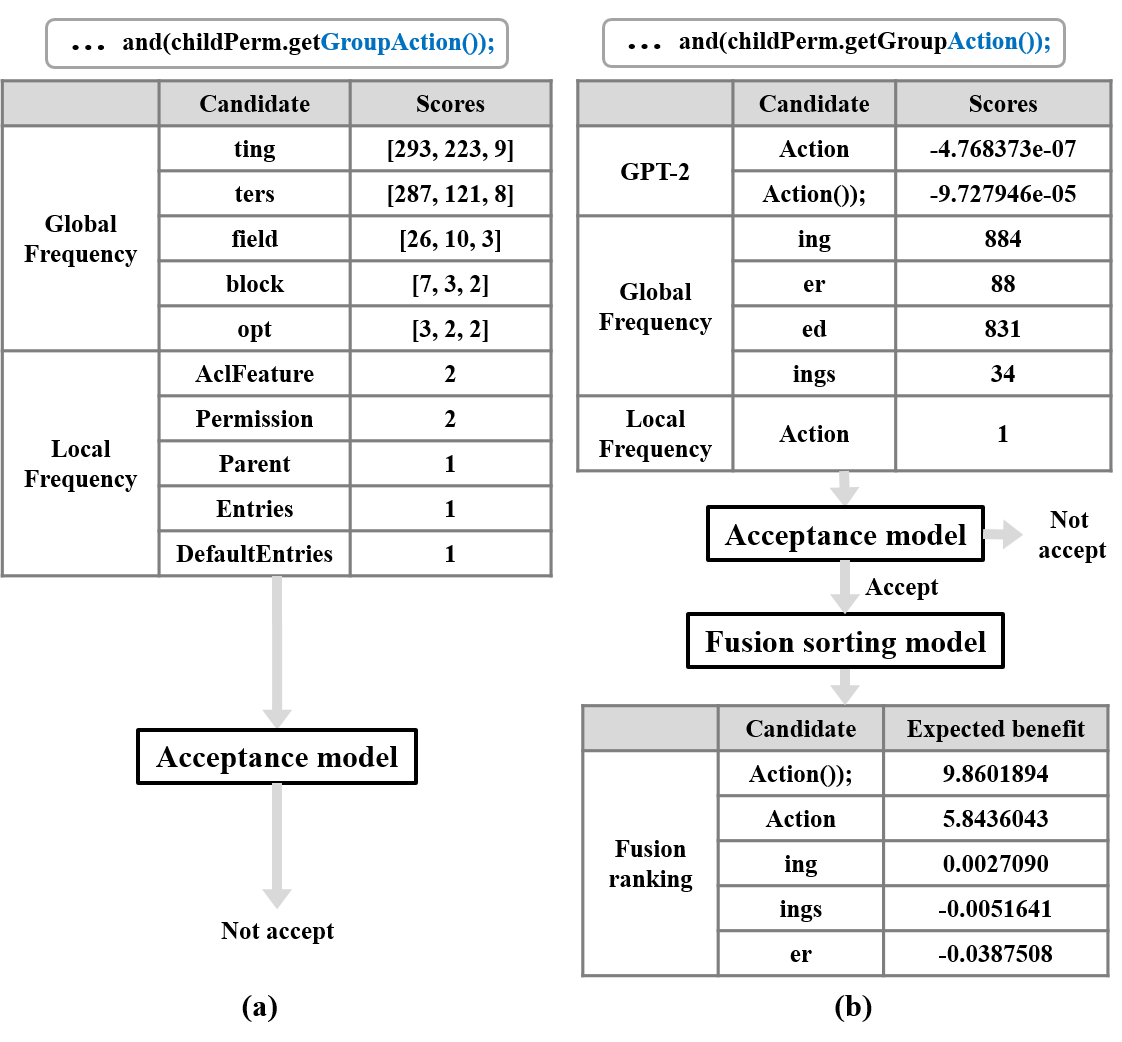}
  \caption{Completion examples with acceptance and fusion ranking models. Example (a) shows a scenario that no candidate is correct; example (b) shows a scenario that two candidates are correct, the longer one is ranked at the top position}
  \label{fig:example2}
\vspace{-0.5cm}
\end{figure}

\section{Dataset} 
\label{section:Dataset}

To generate the dataset, we take the Java-small dataset of Alon et al.\cite{alon2019code2vec}, which is a reselect of the dataset of Allamanis et al.\cite{allamanis2016convolutional}. It involves the most popular eleven Github projects, such as Cassandra, ElasticSearch, Gradle, etc. There are about 96.5k Java source files in total. It removes the overridden, abstract or class constructor methods in every Java file, and rewrite the names of camelCase into snake\_case. We further conduct the following filters:
\begin{itemize}
\item Remove the files which name contains the word “test”;
\item Remove the methods named “toString”, “equals”, “finalize”, and “clone”;
\item Remove the methods longer than twenty lines as they are normally used for configuration or initialization.
\end{itemize}
Finally, the dataset contains 89.3k/1.6k/3.5k training/simulation/test files. The training dataset is used to train each integrated code completion strategy. The simulation dataset is used to run the simulation scenario and collect completion results from each strategy to train the acceptance and fusion ranking models. Finally, the test dataset is used for the result evaluation.

\section{Simulation Data Generation}
\label{section:Simulation data Generation}
To optimize the completion list from the user experience perspective, we use a simulation method to collect the candidate sets of each completion strategy. Then automatically label validated candidates according to the actual code content at completion position. Considering that the simulation process is relatively time-consuming, we speed up the entire data collection process through parallelism.

\subsection{Data generation}

In actual application, the more diverse strategies are integrated, the more significant effect of the model ensemble can be observed. We select three recall strategies for optimization:

\begin{itemize}
\item Global Frequency provides the completion of common words when developers define new variables or functions. 
\item Local Frequency captures the local repetition of the variables, classes and methods, which can infer the corresponding item based on the characters you have typed.
\item GPT-2 styled language model trained from the large codebase can provide intelligent full line code completion.
\end{itemize}

Their implementation details will be described in the experiment setup. And the results section also shows the statistical data of their different characteristics.

When entering a code snippet, for instance, in the black font of the Java file in Fig. \ref{fig:a_location_of_the_java_file}, each strategy will give a completion list. In addition to candidate items themselves, each strategy output scores for its candidates to support the sorting within the strategy. But the meanings of scoring in different strategies are diverse:
\begin{itemize}
\item Global Frequency has three scoring dimensions: word count, the number of occurrences in different documents, and projects involved. These statistics are counted from the entire codebase. Word count is used to sorting candidates within the strategy.
\item Local Frequency only counts the token occurrences of the code before the cursor in the current file.
\item The score of GPT-2 is the cumulative sum of the output of the log-softmax function at each step. The score is decreasing with the increase of the output token sequence length, so the top candidates tend to have a short result.
\end{itemize}

\begin{figure}[tbp]
  \centering
  \setlength{\abovecaptionskip}{-0.2cm}
  \includegraphics[width=0.85\linewidth]{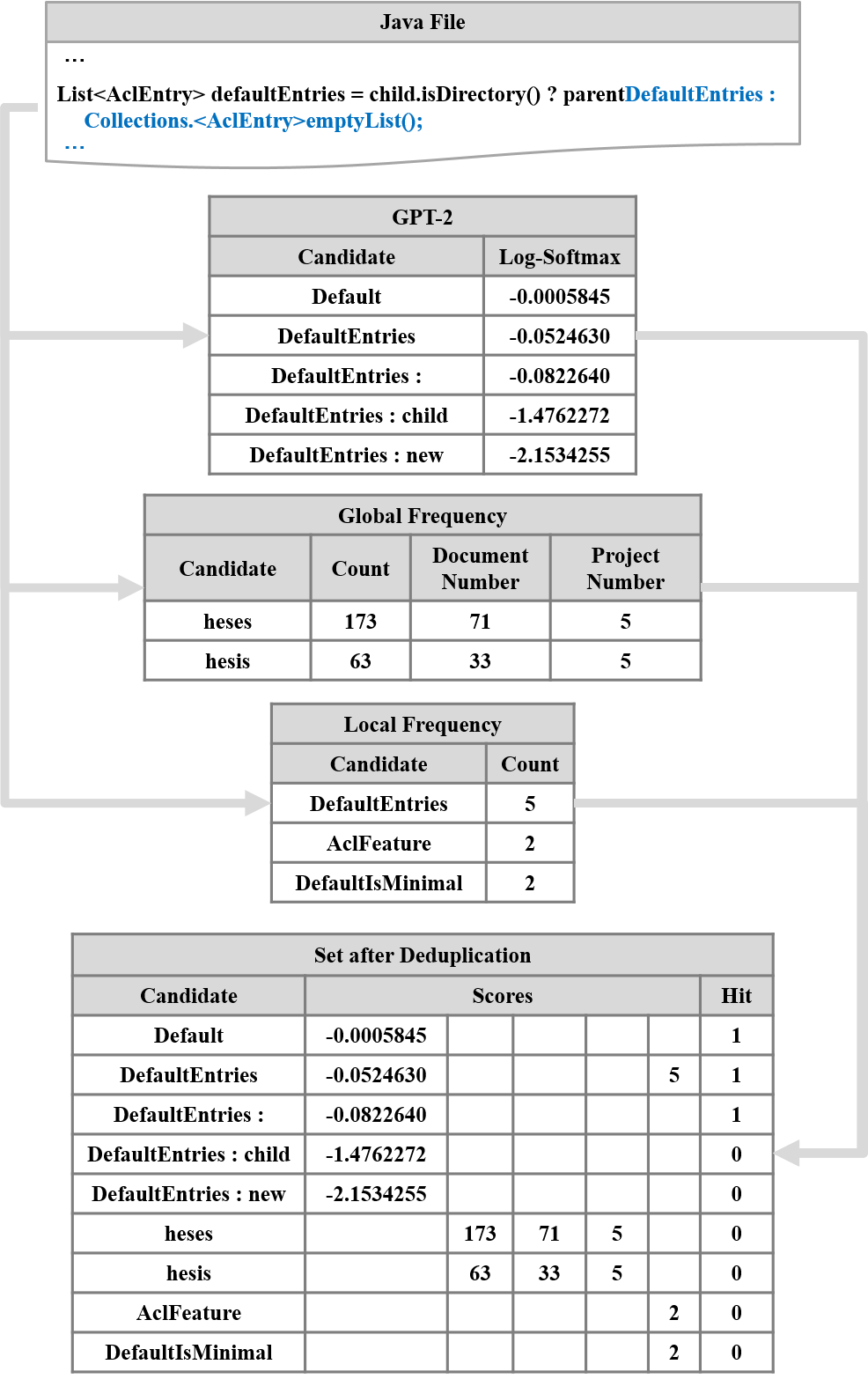}
  \caption{Example simulation result at a certain location. The black font in the Java file is the code fragment before the cursor. The blue font is the code snippet after the cursor, which the target to predict. The three tables in the middle show the original output of the different strategies. The bottom table shows the aggregate completion candidates with a "Hit" column. If one completion candidate matches the target, we set its corresponding value in the "Hit" column as 1, otherwise 0. }    
  \label{fig:a_location_of_the_java_file}
\vspace{-0.5cm}
\end{figure}

When different strategies suggest a candidate at the same time, such as "DefaultEntries" in Fig. \ref{fig:a_location_of_the_java_file}, the scores of it will merge from each dimension for the subsequent tasks.

\subsection{Coding simulation}

It would be great if the online usage data could be collected from the developers when coding. However, it isn't easy to collect the candidate selected by the user in real scenarios and the corresponding context due to factors such as personal privacy. In addition, the cold start is also a problem when system builders want to implement new strategies, as they do not have any data for analysis in advance. So we simulate the developers' coding process, including the typing and the completion selection. Thus we can collect the input code before the cursor, the prediction results from each completion strategy, and the matching labels close to the real scenarios. In addition, because the simulation is offline, the combination of various strategies is possible.

In the process of generating simulation data, we know the actual code context after the cursor, for example, the code snippet in blue font of Java file in Fig. \ref{fig:a_location_of_the_java_file}. Therefore, we can label each candidate by the prefix matching method. The "Hit" column in the bottom table of Fig. \ref{fig:a_location_of_the_java_file} marks the correct completion results as 1 and keeps the others as 0. 

For each file in our simulation dataset, we conduct the simulation by moving the cursor from the beginning to the end by one character per step. We generate one sample at each step. This sample includes the input code snippet (the code context before the cursor) and its corresponding set candidates from integrated strategies. Thus, the number of simulation samples generated from a file is equal to the length of the file. 

\subsection{Critical position and Non-critical position}
\label{subsection:Critical position and Non-critical position}
To emphasize the samples that most likely to appear in real scenarios, we define critical/non-critical positions in the code file and classify the simulation samples by the types of their corresponding trigger positions. Critical position means the position where the code completion request is more likely to be triggered. We filter the simulation data samples by dynamically skipping the non-critical positions where their corresponding characters are generated by the completion tool. The remaining positions are marked as critical positions as they are where the code completion is triggered in real scenarios. Fig. \ref{fig:critical_position} presents the critical position with underlined characters. It should be noted that the critical positions can vary when applying different completion strategies as they depend on the completion results at each position. In this work, we mainly use simulation samples at the critical positions for further analysis as they are more representative.
In our simulation, the proportion of critical positions is about 23.17\%. We also found that the completion accuracy is as high as 90\% when considering all positions, while it drops to 55\% when only considering critical positions. It indicates that the chosen critical positions are crucial for the overall code completion evaluation.

\begin{figure}[tbp]
  \centering
  \setlength{\abovecaptionskip}{-0.2cm}
  \includegraphics[width=0.9\linewidth]{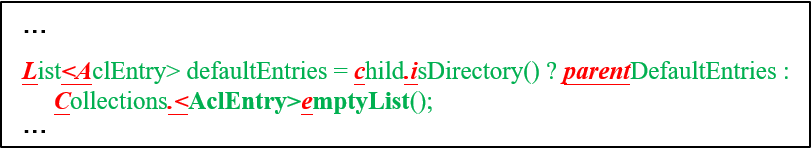}
  \caption{Training samples at critical position. The red underlined characters are the critical positions.}
  \label{fig:critical_position}
\vspace{-0.5cm}
\end{figure}

\subsection{Speed up through parallelism}
Since the time to simulate a single file is at the level of minutes or more, we deploy each completion strategy as services that can be scalable. And the interaction between master and worker is asynchronous. Master will take out an idle worker from the pools and then send a file-level simulation task request. The choice of server cluster size is generally related to the number of code files involved in the simulation and the efficiency of the strategy to generate the completion list. We used 20 EC2s from cloud service with the machine specifications of s3.xlarge.2 to generate results for the simulation and test dataset. To simulate the data for the code files mentioned in Section \ref{section:Dataset}, it took about 2.5 days with the parallel mechanism, significantly shorter than 50 days with only one worker.

\section{Acceptance Model}
\label{section:Acceptance of completion list}
In general, the increase in the number of integrated completion strategies will raise the trigger probability of code completion. Although the recall is improved, it also produces more false-positive completion results. These invalid results will superimpose additional costs that reduce development efficiency. So we design a module to judge whether or not to accept the completion list by evaluating its confidence. If the module accepts the completion result, the result will display at the corresponding position; otherwise, the module will block the completion result.

\subsection{Module design with candidate-set level samples}
It is time-consuming to manually design a set of acceptance rules for this module based on expert experience. In addition, if the integrated completion strategies are changed, it is difficult to update the bespoke rules in time. Therefore, we design an acceptance model, adopting the method of binary classification modelling. This model is trained by the simulation samples at critical positions (Section \ref{subsection:Critical position and Non-critical position}) to better fit with coder experience. When the cumulative of "hit" columns (Fig. \ref{fig:a_location_of_the_java_file}) in a candidate set is greater than 0, label this completion action as a positive sample; otherwise it is negative.

\subsection{Model algorithm}
To construct this binary classification model, we use Autogluon package\cite{erickson2020autogluon} in the pipeline of data preprocessing, model selection and hyperparameter tuning. Autogluon can train models with commonly used classification algorithms. Moreover, it can construct ensemble models with multiple algorithms and different hierarchical structures. Table \ref{tab:Autogluon_result} shows the offline evaluation results of six classification models and one ensemble model trained by Autogluon with our training dataset for the acceptance model. In addition, the table presents the classification accuracy, prediction time and training time of each model. Considering the time constraint in the real-time code completion scenario, we finally chose the LightGBM algorithm. LightGBM is a distributed gradient boosting framework based on the decision tree algorithm, providing fast and accurate prediction with a low memory footprint.  To further improve the online inference performance, we apply the native LightGBM\cite{ke2017lightgbm} in our online service and reduce the prediction time to less than 50ms.

\begin{table}[tbp]
\setlength{\belowcaptionskip}{-0.1cm}
\centering
\caption{Offline evaluation results of different classification models trained by Autogluon}
\resizebox{0.9\linewidth}{!}{
\begin{tabular}{|c|c|c|c|}
\hline
\textbf{Model} & \textbf{Accuracy} & \textbf{Pred Time (s)} & \textbf{Train Time (s)}\\
\hline
Weighted\_ensemble & 93.49\% & 3.331 & 3881\\
\hline
RandomForestClassifier & 92.98\% & 0.246 & 77\\
\hline
ExtraTreesClassifier & 92.45\% & 0.446 & 62\\
\hline
CatboostClassifier & 89.24\% & 0.114 & 23\\
\hline
NeuralNetClassifier & 91.02\% & 1.836 & 3586\\
\hline
LightGBMClassifier & 90.94\% & 0.069 & 5\\
\hline
KNeighborsClassifier & 90.36\% & 4.643 & 488\\
\hline
\end{tabular}}
\label{tab:Autogluon_result}
\vspace{-0.5cm}
\end{table}

\subsection{Feature importance for acceptance model}

The features we used are from two sources: the code context to capture position and structure features from the existing code and the completion results to capture features from each code completion model. 

\subsubsection{Code context features}
Extract context features from the code before the position to be completed, mainly from the following aspects:
\begin{itemize}
\item Position: line number, token number of the line, etc. 
\item Uncompleted token: prefix length, capitalized token, etc.
\item Adjacent tokens: last token, last symbol, etc.
\end{itemize}
\subsubsection{Completion result features}
Collect features from completion results of each integrated strategy. Some additional features are generated by aggregating the results from all strategies. 
\begin{itemize}
\item Feature from each strategy: candidate number, score of each candidate, candidate length, etc. 
\item Aggregate features: the times of simultaneous occurrences of a candidate in multiple strategies, etc.
\end{itemize}

Fig. \ref{fig:acceptance_model_feature_importants} presents the top fifteen critical features in the LightGBM model, the importance score denotes the number of times the feature used in this model. It can be noticed that the top three important features come from the current line code context. The adjacent token and symbol play an essential role as they are most related to the code to be predicted. On the other hand, eleven out of these fifteen features are from the completion results, especially from the length and scores of the top two candidates of each strategy. In addition, one aggregate feature, the times of simultaneous occurrences of a candidate in multiple strategies, is presented as the fourteenth important feature. 

\begin{figure}[tbp]
  \centering
  \setlength{\abovecaptionskip}{-0.1cm}
  \includegraphics[width=0.9\linewidth]{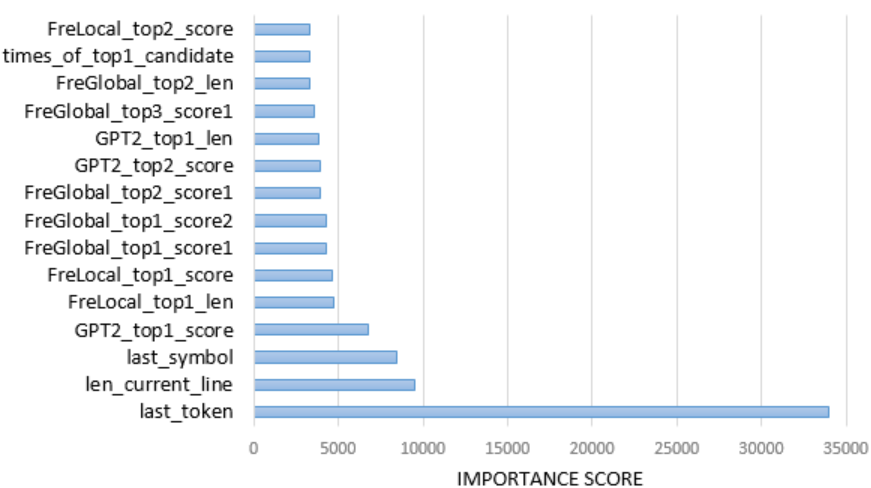}
  \caption{Feature importance of the acceptance model}
  \label{fig:acceptance_model_feature_importants}
\vspace{-0.5cm}
\end{figure}

\section{Fusion Ranking Model}
\label{section:Fusion ranking method}
When a correct candidate does not present at the top of the completion list, the developer has to increase the checking time to scan all the invalid candidates above it and increase the tap times to select it. Thus, the position of the correct candidate in the list significantly affects the coding efficiency. When the code completion system has multiple strategies, the candidate scores given by different strategies are usually on various scales, and sometimes one strategy might have a multi-dimensional score. To display the candidates from all integrated strategies in an ordered list, adjusting the candidates' priority is needed.  

\subsection{Module design with candidate level samples}
Unlike the candidate-set level sample granularity of the acceptance module, we extract candidate level samples from the simulation data at critical positions for this fusion ranking module. In this manner, one completion candidate is treated as a sample. Thus, the total number of samples will expand dramatically. Assuming the acceptance model has blocked the completion without any correct candidate, we only select the samples from the completion list with the correct candidate, which also solves the problem of the unbalanced samples. 

Similar to the acceptance model's sample features, each candidate sample in the ranking model has code context features and features related to the candidate itself. The goal of the ranking model is to give a score as a unified measurement index to each candidate sample from different strategies. This score then ranks all the candidates in a completion list.

\subsection{Model algorithm}
\label{subsection:sorting_algorithm}
To rank candidates from multi strategies, we have explored the following two methods:
\subsubsection{Normalized Ranking}The first method we considered is to normalize the candidate scores from each strategy by the StandardScaler function in the scikit-learn package\cite{scikit-learn}. It scales each input score separately by subtracting the mean and dividing by the standard deviation to shift the distribution to a mean of zero and a standard deviation. Then we rank all candidates by this normalized score. However, the top-ranking candidates tend to be more accurate and shorter in length with this method as the candidate length factor is not considered here.
\subsubsection{Fusion Ranking}It is not easy to introduce a length factor\cite{svyatkovskiy2020intellicode,wu2016google} into the ranking score and to achieve a compromise between the two indicators. We innovatively introduce the concept of expected benefit together with a regression model to deal with this problem. In this regression model, the input is the feature vector of a candidate, and the output is the expected completion length of this candidate. More specifically, for the correct candidate, use its length as the model's output; for the wrong candidate, set the output as 0 directly. The physical meaning of this method represents the potential benefits of choosing this candidate. If the candidate is wrong, there is no help to improve the coding efficiency, and when it turns out to be correct, the longer the candidate's length will bring more significant benefit.

We still choose the LightGBM algorithm to train this regression model with simulation data. The output of the regression model inherently considers both the correct probability and the length of the candidate. Thus, our fusion ranking module takes the output of the regression model as the final score and ranks candidates by this score. Fig. \ref{fig:an_example_of_score_module} is an example of the fusion ranking module. It can be seen that the output score from the regression model believes that compared with candidates 4 or 5, candidate 1 has a greater profit of giving longer completion compared to its risk of accuracy. In contrast, the risk of accuracy overrides the length benefits for candidates 2 and 3. In this manner, the fusion ranking module gives the top-ranking place to the correct candidate with the longest length.

\begin{figure}[tbp]
  \centering
  \setlength{\abovecaptionskip}{-0.1cm}
  \includegraphics[width=0.7\linewidth]{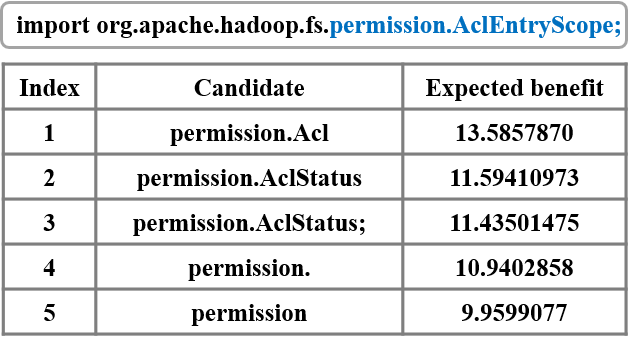}
  \caption{Example of fusion ranking module}
  \label{fig:an_example_of_score_module}
\vspace{-0.5cm}
\end{figure}

\section{Evaluation Metrics Considering Hidden Cost}
\label{section:Evaluation Method}

In this section, we provide a novel comprehensive evaluation method. Unlike the simple accuracy metrics used in the previous code completion research, our metric takes into account the benefits from the code completion, meanwhile considers the hidden cost of it. Thus, this metric is closer to the real coder experience.

\subsection{Common metrics}

The following performance indicators are commonly used to evaluate code completion models and algorithms: in academia,  Accuracy of top K completion results (Accuracy@K)\cite{boyd2012accuracy} and Mean Reciprocal Rank (MRR)\cite{radev2002evaluating, voorhees1999trec}
are representative evaluation metrics. 
On the other hand, the keystroke is commonly used in industry to evaluate code completion systems, such as the well-known code auxiliary tools aiXcoder\cite{aiXcoder}, Kite\cite{Kite-web} and TabNine\cite{Tabnine}.

In this paper, we use Accuracy@K as the common metric for evaluating the effect of code completion. The definition formula is as follows:

\begin{equation}
\label{eqn:Acc}
    Accuracy@K = \frac{\sum_{i \in W}isHit_{(i,K)}}{|W|}
\end{equation}

$W$ represents the set of locations that have completion results in the test code file. For each location $i$ in $W$, the value of $isHit_{(i,K)}$ is 1 when the correct completion result is in the top $K$ results of the completion list. Otherwise, the value is 0.

\subsection{Benefit-Cost Ratio}

In the industry, when propagating the effect of the completion plug-in, it is more inclined to use the concept of the keystrokes, which is the number of times that a human needs to type to complete a document. Therefore, if the code completion tool can significantly reduce the keystrokes, coding can be more efficient. 

Although the accuracy index can compare the performance of some models to some extent, it is incapable of evaluating models with different types of output. For example, the model with single token output generally has higher accuracy than the model with multi-token output. On the other hand, the keystrokes benefit reflects the extent to which code completion tools improve user efficiency. However, it ignores the hidden cost of browsing the completion list. To solve these issues, we propose our comprehensive evaluation metric, Benefit-Cost Ratio(BCR), for the code completion task.

\subsubsection{Benefit}
First we define the $Benefit$, which is the number of keystrokes reduced by using the code completion tool, with the following formula:

\begin{equation}
\label{eqn:Benefit}
    Benefit = N_{ori} - N_{cc}
\end{equation}

where $N_{ori}$ is the number of characters in the test file, more specifically, it means the keystrokes without using the code completion tool; $N_{cc}$ represents the number of actual keystrokes with the usage of code completion tool. It should be noted that $N_{cc}$ defined here does not involve the taps used to selecting the right answer in the completion list as this will be counted in the hidden cost.

\subsubsection{Hidden Cost}
Code completion tools not only improve development efficiency but also bring inevitable usage costs to developers. These extra costs, such as the time to browse the completion list and the keystrokes cost to select the correct answer, are often implicit and ignored by previous evaluation metrics. From a practical perspective, we only consider the time to browse the completion list in our work. The costs of the keystrokes to select the correct answer are negligible as they often happen during the browsing procedure and do not need much mental work.  

Assuming developers scan the completion list in a top-down manner and the effect of evaluating each candidate is identical, we can define the browsing cost by the following rules: 
\begin{itemize} 
\item if the correct answer is in the completion list, the browsing cost is equal to the ranking of the correct answer in the completion list as the developer would evaluate all the invalid answer above it. In case of more than one correct answers appeared in the completion list, we use the position of the longest correct answer; 
\item if none result in the completion list is correct, the browsing cost is equal to the length of the completion list, as the developer would have to evaluate all the invalid candidates in the list. 
\end{itemize}
Thus, we define the $HiddenCost$ with Eq. (\ref{eqn:hidden_cost}):

\begin{equation}
\label{eqn:hidden_cost}
    HiddenCost = \sum_{i \in Hitset}HitPos_{i} + \sum_{j \in (W - Hitset)}ListLen_{j}
\end{equation}

where $Hitset$ represents a set of locations in the test files where the correct result is in their completion lists; $W - Hitset$ indicates the set of locations where all the completion results in the completion list are invalid;
The $HitPos_{i}$ indicates the longest correct result position in the completion list at location $i$ of the test file; $ListLen_{j}$ indicates the length of completion list when there is no correct result in the completion list at location $j$ of the test files.

Based on Eq. (\ref{eqn:Benefit}) and (\ref{eqn:hidden_cost}), BCR can be defined with Eq. (\ref{eqn:BCR}). 
\begin{equation}
\label{eqn:BCR}
\begin{aligned}
    BCR &= \frac{Benefit}{HiddenCost} \\
\end{aligned}
\end{equation}
This equation considers the total benefit brought by the code completion tool together with its hidden cost. The physical meaning of BCR can be roughly regarded as the average saving keystrokes by browsing one candidate in the completion list during the coding process. Taking Fig. \ref{fig:an_example_of_score_module} as an example, the top 1 candidate is correct and has a length of 14. If the developer browses and select this candidate, the browsing cost is 1 and the tap times is 1. It saves 13 keystrokes comparing to manually tap 14 times to complete this token, so the benefit is 13. Thus, the BCR is 13/1 in this case. A higher value of BCR indicates the lower the effort required for the developer and the better the overall effect of the code completion tool.

\section{Experiment Setup} 
\label{section:Experiment Setup}

This section introduces the implementation details of the three strategies (two frequency models and a GPT-2 model) and their corresponding parameter configuration during the experiment. The data set and the evaluation method have been introduced in Section \ref{section:Dataset} and Section \ref{section:Evaluation Method}) respectively. Thus, this section focuses on the setup of the code completion strategies. Each strategy provides up to five candidates to our acceptance and fusion ranking models for the subsequent tasks.  

\subsection{Frequency strategies}

Traditional code completion tools usually provide completion list sorted alphabetically. Such kind of list requires an extra cost to the developer to find the correct result. Sorting by the probability of the candidate occurrences in the historical codebase can improve the relevance of the suggestions, and it can automatically adapt to the changes in the codebase. Thus, we introduce two word-frequency-statistics methods in our experiment. They are integrated as different completion strategies to provide candidates for our acceptance model and fusion sorting model.

\subsubsection{Global Frequency}
When defining variables or functions, developers usually name them with multi-words linked by camel-case or underscore to explain the specific meaning. If we perform code segmentation by token granularity, the long tail effect of the programming language will be more evident than the general natural language. Therefore, we adopt a finer-grained sub-token segmentation method \cite{svyatkovskoy2020fast} to further segment tokens when encountering underscore or camel case naming rules. In addition, we filter out the sub-tokens that only appear in one project, and the length is smaller than 5. As a result, the vocabulary size is dramatically reduced from 600,000 for the token level segmentation to 15,916 after sub-token level segmentation and the filter. Such vocabulary is sufficient to reflect the common words used in the coding process. Besides the keywords and basic types, the most frequently used tokens are (1) Terms that often appear in Java file header comments such as "License", "Version", and "Apache", (2) Words that often appear when importing packages such as "Service", "Manager", and "kernel". It is also noted that "result" has the highest probability as the variable name, and the "append" method has the highest usage rate.

We store the entire vocabulary with a tire tree to accelerate the online inference. For each token, the total number of occurrence in the training set, the number of involved files, and the number of involved projects are recorded as the three dimensions of token scores. Candidates are obtained by searching the tire tree and then sorted according to the total number of occurrences. 

Global Frequency strategy is mainly used when defining new variables, function names, and classes. It can effectively complete the commonly used words in the code file.

\subsubsection{Local Frequency}
The programming language often has distinct localization characteristics\cite{tu2014localness}. When coding, developers may repeatedly call the variable or function defined in the previous code fragment or the custom class under the same project. Therefore, we use the Local Frequency strategy to capture and predict the local characteristics. Local Frequency strategy extracts the token-level vocabularies from the current code file, meanwhile, counts the number of occurrence of each token as the score. During the online inference, these vocabularies are dynamically updated with the written code file. Candidates are obtained by searching the vocabulary directly and then sorted according to the number of occurrences.

As an illustration, we select a representative file ("AclStorage.java") from the testing dataset with 689 tokens and about 8000 characters. Among 689 tokens, 488 tokens have appeared in the code before their location. In other words, the recall rate is 69.91\%. Table \ref{tab:token_occurrences} shows the results sorted by the number of token occurrences. It can be seen that the top-ranking tokens repeatedly appear in the code file. In addition, further analysis shows that in the case of the Local Frequency strategy hits, there is about half chance that the correct candidate can be found at the top one position of the completion list of Local Frequency strategy.

\begin{table}[tbp]
\caption{number of token occurrences in a Java file}
\setlength{\belowcaptionskip}{-0.1cm}
\centering
\resizebox{\linewidth}{!}{
\begin{tabular}{|c|c|c|c|c|c|}
\hline
\textbf{Index} & \textbf{Token} & \textbf{Count} & \textbf{Index} & \textbf{Token} & \textbf{Count}\\
\hline
1 & AclEntry & 30 & 6 & FsPermission & 15\\
\hline
2 & accessEntries & 27 & 7 & import & 14\\
\hline
3 & List & 20 & 8 & permission & 14\\
\hline
4 & inode & 19 & 9 & featureEntries & 13\\
\hline
5 & if & 19 & 10 & new & 13\\
\hline
\end{tabular}}
\label{tab:token_occurrences}
\end{table}

\subsection{Sentence-level language model}
With their self-attention and paralleled processing mechanism, the newly designed transformers present superiority over RNNs in modelling the long-term dependency and the inference speed. GPT-2\cite{radford2018improving}, as a transformers styled pre-train language model, has demonstrate its strong ability in both text and code generation area\cite{svyatkovskiy2020intellicode}. Thus, we use GPT-2 styled model as the sentence-level language model in our research.

\subsubsection{Model structure}
GPT-2 consists of a multi-layer transformer decoder stack, which maps input token embedding and position embedding into an output vector. The output vector is then multiplied with the token embedding matrix and forward into a log-softmax function to calculate the prediction score for each vocabulary token. We train our GPT-2 model from scratch with the training dataset.

Considering the trade-off between accuracy and performance, we select a small GPT-2 124M model and apply a minor modification to build our own GPT-2 model for the code completion task. The maximum sequence length is reduced from 1024 to 256, and the vocabulary size is reduced from 50257 to 30000. Final model hyper-parameters are shown in Table \ref{tab:GPT_paras} with 108M parameters in total.

\begin{table}[tbp]
  \caption{GPT-2 model architecture hyper-parameters}
  \setlength{\belowcaptionskip}{-0.1cm}
    \centering
  \resizebox{0.8\linewidth}{!}{
  \begin{tabular}{|l|l|l|}
    \hline
    \textbf{Hyper-parameter} & \textbf{Explanation} & \textbf{Value}\\
    \hline
    n\_layer & Number of transformer layers & 12\\
    \hline
    n\_model & Dimension of hidden states & 768\\
    \hline
    n\_embd & Dimension of embedding & 768\\
    \hline
    n\_head & Number of attention heads & 12\\
    \hline
    n\_positions & Max code token length & 256\\
    \hline
    p\_drop & Drop probability & 0.1\\
    \hline
    vocab\_size & Vocabulary size & 30000\\
  \hline
\end{tabular}}
\label{tab:GPT_paras}
\vspace{-0.5cm}
\end{table}

\subsubsection{Preprocessing}

We apply the following preprocessing steps to the source code, including:
\begin{itemize}
  \item Partially remove comments, only keep comment lines before the function definition and docstring below the function definition;
  \item Remove non-English letters and symbols; 
  \item Replace long string/number with special placeholders;
\end{itemize}

\subsubsection{Tokenizer}

The tokenizer is used to encode a code literal into a sequence of tokens before feeding it into GPT-2 model and decode the model output back to a code literal. We use the same tokenization method, Byte-Pair Encoding (BPE), with the original GPT2 model. It is an unsupervised tokenizer, which recursively replaces the most frequently occurring pair of Unicode characters with a new character in the vocabulary. We retrain the BPE tokenizer from scratch using our source code data and set the vocabulary size as 30000. Since this customized tokenizer provides a better fit with source code, it can consider more extended source code than the original GPT-2 tokenizer with the same number of tokens.

\subsubsection{Inference}

\begin{figure*}[bhp]
\vspace{-0.5cm}
  \centering
  \setlength{\abovecaptionskip}{-0.1cm}
  \includegraphics[width=0.8\linewidth]{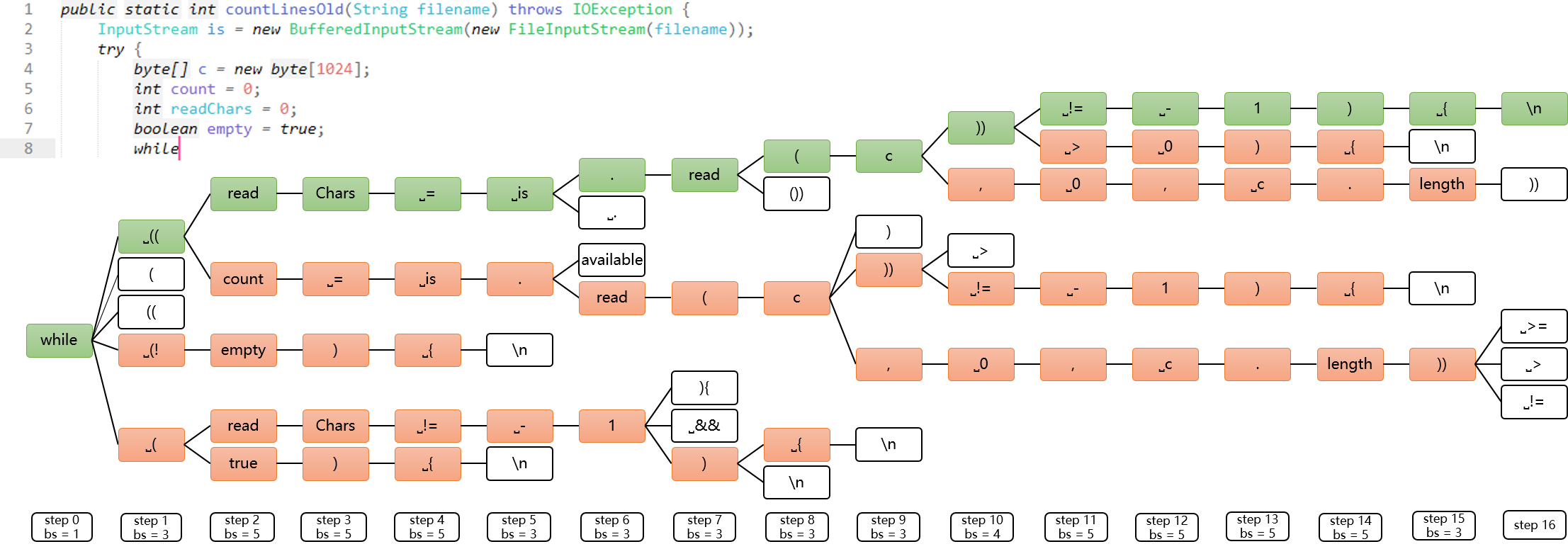}
  \caption{GPT-2 beam-search. The green branch is the one with highest aggregate probability; the white blocks are the ones match the termination criterion}
  \label{fig:beam_search}
\vspace{-0.5cm}
\end{figure*}

The maximum online inference time for GPT-2 model is set as 200ms to ensure displaying completion results to the developer fluently. We apply the beam-search method (Fig. \ref{fig:beam_search}) in the GPT-2 model inference procedure to generate the code completion result. Beam-search reduces the risk of missing high probability token sequences by keeping the top $k$ ($k$ is the beam size) highest aggregate probability sequences at each step. To reduce the redundant calculation and accelerate the inference during the beam-search, we apply the following output sequence termination criterion to adjust the beam size dynamically at each step:

\begin{itemize}
  \item if its aggregate score is lower than a score threshold $t$. 
  \item if it has end-of-line token ("$\backslash$n").
  \item if it has an annotation symbol.
  \item if it runs into closed loop.
  \item if it reaches the maximum inference step.
\end{itemize}

Fig. \ref{fig:beam_search} illustrates one of our beam-search example with beam size $k$ and score threshold $t$ being set as 5 and -3 respectively. At step 0, the top 5 score candidates are selected and combined as a batch for the step 1 inference. At step 1, we also take the top 5 candidates with the highest aggregate score, in which only three candidates have an aggregate score higher than the threshold $t$. Thus, only three candidates are selected for the next step inference, which means the batch size of step 1 input is 3. The rest inference steps are conducted in the same manner until our criterion terminate all beam search process. Thus, instead of doing beam-search with fixed batch size inference at each step, we dynamically adjust the batch size according to the previous step result to reduce the inference cost. 

Given the increment sequence typing nature of the coding process, we also implement two caching mechanisms in the inference stage. Firstly, we cache the input sequence with its attention keys and values of the GPT-2 blocks. If a new inference request is received, we only calculate the keys and values for the additional transformer block if the input sequence partially matches the items in the cache. This caching speeds up inference by up to 40\%. Secondly, we cache the inference sequence results for the current line. When the beam-search procedure cannot finish in the time constraint (200ms), we return the results from the processed step. Meanwhile, we keep the beam-search running in the background until it reaches the termination criterion. The untapped results are saved in cache and used if a new inference with a matching prefix is received.

\section{Result}
\label{section:Result}

This section conducts a series of experiments of three chosen code completion strategies with our acceptance model and fusion sorting model and presents a comprehensive analysis based on the experimental results. The data used for the evaluation is generated from the test dataset introduced in Section \ref{section:Dataset} and \ref{section:Simulation data Generation}.

\subsection{Single strategy evaluation}
Table \ref{tab:Characteristics} shows the comparison of the characteristics of different strategies from the following aspects.
\begin{itemize}
\item Occurrence rate: the proportion of non-empty completion list.
\item Hit position 90\% quantile: when the completion list has the right candidate, collect the positions of the longest hit candidate. Sort the collected position in ascending order and take the value of 90th quantile to avoid discrete points.
\item Prefix length 90\% quantile: prefix length means how many characters of a token you need to tap before the correct result appears in the completion list. We also use the value of the 90th quantile.
\item Completeness: the proportion of the chosen candidate is a full token. 

\item Accuracy@K: as described in Eq. (\ref{eqn:Acc}).
\end{itemize}

\begin{table*}[tbp]
\setlength{\belowcaptionskip}{-0.1cm}
\centering
\caption{Characteristics of different code completion strategies}
\resizebox{\linewidth}{!}{
\begin{tabular}{|c|c|c|c|c|c|c|c|}
\hline
\textbf{Strategy} & \textbf{Occurrence rate} & \textbf{Hit position 90\% quantile} & \textbf{Prefix length 90\% quantile} & \textbf{Completeness} & \textbf{Accuracy@1} & \textbf{Accuracy@5}\\
\hline
Global Frequency & \textbf{47.33}\% & \textbf{TOP2} & 12  & 65.80\% & 9.74\% & 11.18\%\\
\hline
Local Frequency & 12.48\% & TOP4 & \textbf{2}& \textbf{95.03\%} & 31.82\% & 52.67\%\\
\hline
GPT-2 & 8.20\% & TOP5 & 9 & 88.78\% & \textbf{58.96\%} & \textbf{62.79\%}\\
\hline
\end{tabular}}
\label{tab:Characteristics}
\vspace{-0.5cm}
\end{table*}

\subsubsection{Global Frequency}
Global Frequency is able to provide results at almost 50\% of the locations while has the lowest accuracy among these three strategies. The long requirement of the prefix length helps the Global Frequency strategy hit the correct result at a relatively top position as the extended prefix limits the number of candidates with solid constraint. When there is a hit item in the completion list, the correct completion result is ranked in the top 2 positions in 90\% of the cases. In addition, since the Global Frequency strategy segments the code with sub-token level granularity, its completeness is lower than the other two strategies. Most of the correct results of the Global Frequency strategy are the keywords, common types, and the customary name of variables and methods.

\subsubsection{Local Frequency}
Local Frequency represents the localization characteristic of the program coding. Table \ref{tab:Characteristics} shows that, although the occurrence rate of the Local Frequency model is lower than that of the Global Frequency model, an input with only two prefix characters can help the Local Frequency strategy to find the correct completion result in 90\% of the cases. It also has the highest completeness, 95\% of the time, Local Frequency Strategy can complete an entire token.

\subsubsection{GPT-2}
The completion list of GPT-2 has the highest Accuracy@1 and Accuracy@5, with Accuracy@5 reaching 62.79\% compared to 52.67\% and 11.18\% for the Local and Global Frequency strategies, respectively. The occurrence rate is relatively low as we manually limit the trigger condition of the GPT-2 strategy. The ranking method for GPT-2 itself is to sort by the log-softmax score of each candidate, while the shorter candidate generally has a higher log-softmax score. Our fusion ranking model will amend this improper ranking method. In addition, by blocking the low-confidence completion list with our acceptance model, the accuracy can be further optimized.

\subsection{Benefit-Cost Ratio (BCR)}
BCR is the bespoke evaluation metric (defined in Section \ref{section:Evaluation Method}) we proposed for the code completion task. It considers both the benefit and the hidden cost when using the code completion tool in a real scenario. The last row ("Acceptance+Fusion Ranking") of Table \ref{tab:metrics} presents the Accuracy@k and BCR results of our code completion scheme with the acceptance model and fusion ranking model. Our code completion scheme achieves a BCR value of 3.65, which is higher than any single strategy. It means that the developer can averagely save 3.65 keystrokes by browsing one candidate in the completion list when using our code completion scheme.      

It is noted that the BCR of the Global Frequency strategy is very poor. Such performance is consistent with the actual experience. The high trigger and low accuracy completions from the Global Frequency are dazzling. The other interesting finding is that the BCR value of our completion scheme (3.65) is higher than the sum of the BCR value of the three integrated strategies themselves (3.34), indicating that our completion scheme can draw merits and offset defects from each integrated strategy. 

\begin{table}[tbp]
\setlength{\belowcaptionskip}{-0.1cm}
\centering
\caption{Experimental result of the evaluation metrics}
\resizebox{\linewidth}{!}{
\begin{tabular}{|c|c|c|c|c|c|c|c|}
\hline
\textbf{Strategy} & \textbf{Accuracy@1} & \textbf{Accuracy@5} & \textbf{Benefit-Cost Ratio}\\
\hline
Global Frequency & 9.74\% & 11.18\% & 0.09\\
\hline
Local Frequency & 31.82\% & 52.67\% & 1.11\\
\hline
GPT-2 & 58.96\% & 62.79\% & 2.14\\
\hline
Acceptance+Fusion Ranking & \textbf{62.61\%} & \textbf{82.55\%} & \textbf{3.65}\\
\hline
\end{tabular}}
\label{tab:metrics}
\end{table}

Fig. \ref{fig:a_spectrum_example} shows the completion effect of a code snippet in a test file. The abscissa of each subgraph represents the locations of successive code characters in the code snippet. The height of the histogram represents the coding cost of the current position, and the cost value ranges from 0 to 6. The number 0 means no need to perform any action at this position as this character is completed by the previous position. Value 1 may mean that this character is manually tapped, or the top 1 candidate in the completion list is hit. Value 2 to 5 are similar to value 1, and they are just different in the order of the hit candidate position in the completion list. Value 6 means that there is no correct option in the completion list. The developer has to browse five invalid candidates and manually type this character. It can be seen that our completion scheme with the acceptance model and the fusion sorting model can significantly reduce the cost when writing this code snippet.

\begin{figure}[tbp]
  \centering
  \setlength{\abovecaptionskip}{-0.1cm}
  \includegraphics[width=\linewidth]{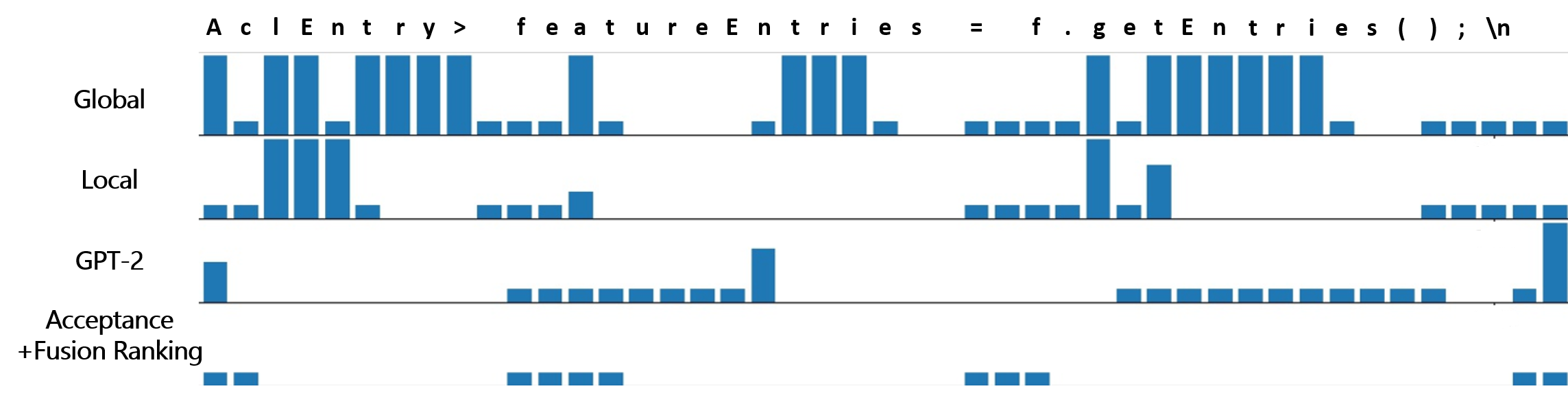}
  \caption{The real cost to write a code snippet}
  \label{fig:a_spectrum_example}
\vspace{-0.5cm}
\end{figure}

\subsection{Ablation study}

Since our proposed code completion scheme involves two subtasks, we conduct a set of comparative experiments to evaluate the effect of each task. The results are presented in Table \ref{tab:ablation_study}. Following strategies are used in this study:
\begin{itemize}
  \item Normalized Ranking: a baseline ranking method introduced in Section \ref{subsection:sorting_algorithm}
  \item Fusion Ranking: only use fusion ranking model
  \item Acceptance+Normalized Ranking: use acceptance model and Normalized Ranking method
  \item Acceptance+Fusion Ranking: use acceptance model and fusion ranking model
\end{itemize}

From $Fusing Ranking$ to $Acceptance+Fusion Ranking$, there is a dramatic improvement in Accuracy@K and BCR metrics, which explains the function of the acceptance model. To verify the effect of the fusion ranking module, we compare $Acceptance+Normalized Ranking$ with $Acceptance+Fusion Ranking$. By replacing the Normalized Ranking with Fusion Ranking, BCR increases from 2.14 to 3.65, which shows that the fusion ranking model also plays an essential role in our code completion scheme.

\begin{table}[tbp]
\caption{Comparison results of different combinations}
\setlength{\belowcaptionskip}{-0.1cm}
\centering
\resizebox{\linewidth}{!}{
\begin{tabular}{|c|c|c|c|c|c|}
\hline
    \textbf{Strategy} & \textbf{Accuracy@1} & \textbf{Accuracy@5} & \textbf{Benefit-Cost Ratio} & \textbf{Invalid list}\\
\hline
    Normalized Ranking & 34.81\% & 44.91\% & 1.20& 55.09\%\\
\hline
    Fusion Ranking & 35.86\% & 50.35\% & 1.57 & 49.65\%\\
\hline
    Acceptance+Normalized Ranking & 57.31\% & 71.80\% & 2.14 & 28.20\%\\
\hline
    Acceptance+Fusion Ranking & \textbf{62.61\% (27.80\%↑)} & \textbf{82.55\% (37.64\%↑)} & \textbf{3.65} & \textbf{17.44\%}\\
\hline
\end{tabular}}
\label{tab:ablation_study}
\vspace{-0.5cm}
\end{table}

We also count the probability of the invalid completion lists with the value shown in the $Invalid\ list$ column. If the acceptance model is not integrated, the likelihood of invalid triggers is nearly 50\%. Such frequently display of the false result increases the expense of declined user experience. After using the acceptance model, the invalid trigger rate reduces significantly. Combining with the fusion ranking model, the probability of $Invalid\ list$ is further reduced to 17.44\%. 

Compared with the baseline strategy "Normalized Ranking",  Accuracy@1 and Accuracy@5 of "Acceptance+Fusion Ranking" are increased by 27.80\% and 37.64\%, respectively.

\section{Conclusion} 
\label{section:Conclusion}

We have introduced and deployed a code completion scheme capable of integrating the results from multiple completion strategies. Within this scheme, we introduced two models: the acceptance model and the fusion ranking model. The acceptance model uses features extracted from the code context and the results from different code completion strategies to predict whether a correct result is in the completion list. It can dynamically control whether to accept the completion results and display them to the developer. The fusion ranking model can automatically identify the priority of the completion results and reorder the candidates provided by different completion strategies. This scheme is flexible in dealing with various code completion strategies, regardless of the type or the length of their completion results. In addition, we have proposed a comprehensive code completion evaluation metric BCR, which considers both the benefit of the keystrokes saving and the cost of completion list browsing. 

We have integrated our code completion scheme with two frequency style models and a GPT-2 style language model to conduct a set of comprehensive experiments. The results prove that our code completion scheme can achieve solid improvements. With the acceptance and fusion ranking models, the proportion of invalid completion list reduces from 55.09\% to 17.44\%. Meanwhile, the TOP1 and TOP5 accuracy increase by 27.80\% and 37.64\%, respectively. The BCR value is 3.65, which indicates the developer can averagely save 3.65 keystrokes by browsing one candidate in the completion list when using our code completion scheme.

To the best of our knowledge, we are the first to optimize the multi-path recall strategies in the code completion task, showing a significant improvement over the single completion strategy. Our exploration may not be comprehensive enough, but we hope to expand new ideas for research in this area. For example, integrating more code completion strategies with different characteristics in the ensemble task may bring more apparent benefits. Trying multi-task joint training or reinforcement learning when optimizing the ensemble model is also an exciting research task. In terms of evaluation methods, we are also expecting to discover more potential influencing factors and combinations. A suitable evaluation method can, in turn, promote the development of algorithms.

\bibliographystyle{unsrt}
\bibliography{main}

\end{document}